\documentclass[amsmath,amssymb]{revtex4-2}

\usepackage{mathtools,bm}
\usepackage{booktabs}
\usepackage{graphicx,xcolor}
\usepackage{newtxtext,newtxmath}
\usepackage{hyperref}
\hypersetup{
  pdftitle={Finite-Difference Zeta Readout of One-Loop Operator Spectra},
  pdfauthor={Keisuke Okamura},
  colorlinks=true,
  linkcolor=blue,
  citecolor=blue,
  urlcolor=blue
}

\DeclareMathOperator{\Tr}{Tr}
\DeclareMathOperator{\tr}{tr}
\DeclareMathOperator{\sech}{sech}
\DeclareMathOperator{\erfc}{erfc}
\DeclareMathOperator{\erf}{erf}
\DeclareMathOperator*{\Res}{Res}

\newcommand{\dd}{\mathop{}\!\mathrm{d}}

\begin{document}

\preprint{arXiv:2604.11460v2}

%*******************************************************************************************
\title{Finite-Difference Zeta Readout of One-Loop Operator Spectra}
%*******************************************************************************************

\author{Keisuke Okamura}
\email{okamura@alumni.lse.ac.uk}
\affiliation{Ministry of Education, Culture, Sports, Science and Technology, Tokyo, Japan}
\date{\today}

%%%%%%%%%%%%%%%%%%%%%%%%%%%%%%%%%%%%%%%%%%%%%%%%%%%%%%%%%%%%%%%
\begin{abstract}
%%%%%%%%%%%%%%%%%%%%%%%%%%%%%%%%%%%%%%%%%%%%%%%%%%%%%%%%%%%%%%%
For a positive elliptic operator $A$, the logarithmic zeta determinant $\ln\det_{\zeta}A=-\zeta_{A}'(0)$ combines UV information encoded by local heat-kernel coefficients with finite contributions determined by the full spectrum.
We introduce a finite-difference zeta readout based on $\zeta_{A}(0)$ and $\zeta_{A}(q-1)$, defining a one-parameter meromorphic family whose node-coalescence limit $q\to 1$ recovers the standard logarithmic zeta determinant.
The parameter $q$ fixes the Mellin evaluation point $s=q-1$, organising genuine poles, regular local special values, generic full-spectrum values, and the logarithmic determinant limit along a common coordinate, while simultaneously determining the spectral weight $\lambda^{-q}$ in the $q$-dependent variational response.
In relative spectral problems, this coordinate distinguishes systems retaining a leading local hierarchy from those in which the entire local power-law hierarchy cancels, illustrated respectively by a reflectionless soliton and a twisted circle.
In four dimensions, the framework recovers the standard local scale response at $q=1$, governed by the heat-kernel coefficient $a_{4}$, whereas at generic regular values of $q$ it retains finite mass-sensitive information beyond the local hierarchy.
The construction thereby provides a unified analytic framework for comparing local UV structure, finite full-spectrum information, variational response, and relative spectral behaviour within fixed operator spectra.
%%%%%%%%%%%%%%%%%%%%%%%%%%%%%%%%%%%%%%%%%%%%%%%%%%%%%%%%%%%%%%%
\end{abstract}

\maketitle

%%%%%%%%%%%%%%%%%%%%%%%%%%%%%%%%%%%%%%%%%%%%%%%%%%%%%%%%%%%%%%%
%%%%%%%%%%%%%%%%%%%%%%%%%%%%%%%%%%%%%%%%%%%%%%%%%%%%%%%%%%%%%%%
\section{Introduction}\label{sec:intro}

The one-loop effective action in quantum field theory is determined by the functional determinant of the quadratic fluctuation operator about a background field.
Since its logarithm is generally divergent, it is defined through regularisation schemes such as cutoff, dimensional, zeta, and heat-kernel regularisation \cite{tHooft72,Hawking77,Vassilevich03}.
The spectral zeta function is related to the heat kernel by the Mellin transform: its poles and residues, together with its regular special values at non-positive integers under appropriate conditions, are determined by the local short-time heat-kernel coefficients \cite{Gilkey95,Vassilevich03}.
By contrast, the finite part of the zeta-regularised determinant depends on the full spectrum and cannot, in general, be reconstructed from any finite set of local heat-kernel coefficients.
Accordingly, the one-loop determinant combines local UV information with global spectral information.
The former governs divergences, counterterms, scale dependence, and quantum anomalies, whereas the latter reflects boundary conditions, topology, holonomy, bound states, and scattering phase shifts.

This coexistence of local and global spectral information is not merely descriptive.
Even operators sharing the same local differential expression may nevertheless possess different finite determinant contributions because of global spectral rearrangements.
Such differences also underlie global and boundary-dependent quantum effects, including the Casimir effect \cite{Milton01}.
In the formulation of relative zeta functions and relative determinants, these spectral differences determine finite relative quantities \cite{Muller98,Yafaev92}.
It is therefore natural to seek a common analytic framework in which local UV structure, finite full-spectrum contributions, and relative spectral differences can be compared on equal footing while remaining consistent with the standard logarithmic determinant.

In the present paper, we formulate a readout rule for a fixed operator spectrum.
Let $A$ be a positive self-adjoint elliptic operator, and let $\nu_{A}(\lambda)$ denote its spectral measure.
A spectral functional may be written formally as
\begin{equation}\label{eq:lam_to_W}
\mathcal{W}[A]=\int_{0}^{\infty}F(\lambda)\dd\nu_{A}(\lambda),
\end{equation}
where the kernel $F$ specifies how the spectrum is read out without modifying the operator itself.
Different choices of $F$ therefore extract different spectral information from the same spectrum.
In this sense, Eq.~\eqref{eq:lam_to_W} is formally related to the Chamseddine--Connes spectral action and its zeta-regularised variants \cite{Chamseddine97,Kurkov15}.
Unlike these constructions, however, the present work does not define a new action but introduces a readout functional acting on the existing spectral zeta function of a fixed one-loop operator.
Appendix~\ref{sec:spectral_operations} further distinguishes this fixed-spectrum readout from Wilsonian coarse graining.

In what follows, we take the logarithmic spectral aggregation appearing in the standard one-loop effective action as the starting point.
The zeta-regularised determinant of such an operator is defined by
\begin{equation}\label{eq:lndet}
\ln\det_{\zeta}A=-\zeta_{A}'(0),
\end{equation}
and its application to one-loop effective actions dates back to the early developments of zeta regularisation \cite{Dowker76,Hawking77}.
In the present work, we introduce a real readout parameter $q$ and replace the derivative evaluation at $s=0$ by a finite difference between the fixed node $s=0$ and the movable node $s=q-1$.
For the finite-difference construction, we take $A$ to be dimensionless and define, wherever both evaluation points are regular,
\begin{equation}\label{eq:qdef}
D_{q}[A]=\frac{\zeta_{A}(q-1)-\zeta_{A}(0)}{1-q}.
\end{equation}
We define the corresponding one-loop readout functional by $\mathcal{W}_{q}[A]=\frac{1}{2}D_{q}[A]$.
As $q\to 1$, the two nodes coalesce, and one obtains $D_{q}[A]\to -\zeta_{A}'(0)$ and $\mathcal{W}_{q}[A]\to -\frac{1}{2}\zeta_{A}'(0)$, thereby recovering the standard logarithmic zeta determinant in Eq.~\eqref{eq:lndet} and the corresponding one-loop effective action.

For suitable classes of elliptic pseudodifferential operators, several related constructions are already available, including the Kontsevich--Vishik canonical trace and determinant \cite{Kontsevich94}, formulae for the Laurent coefficients of the canonical trace of holomorphic families \cite{Paycha07}, weighted traces depending on an auxiliary elliptic weight \cite{Cardona02}, and the residue determinant associated with the Guillemin--Wodzicki residue trace \cite{Scott05}.
By contrast, $D_{q}[A]$ and $\mathcal{W}_{q}[A]$ compare two values of the existing spectral zeta function for a fixed operator without defining a new trace or determinant.
At the algebraic level, their finite-dimensional form involves the power trace $\Tr A^{1-q}$ and is therefore analogous to the power-trace structures appearing in R\'{e}nyi-type quantities \cite{Renyi61}.
In the present construction, however, $q$ shifts the Mellin evaluation point of a fixed one-loop spectrum without introducing either a density operator or a replica geometry.

The central structural feature of the finite-difference family is the common role played by the parameter $q$.
As shown in Section~\ref{sec:readout}, $q$ simultaneously fixes the Mellin evaluation point $s=q-1$ and, in the $q$-dependent component of the operator variation, the response $A^{-q}\delta A$, whose eigenvalue weight is $\lambda^{-q}$.
The Mellin readout and its variational response are therefore two aspects of a single readout coordinate, which also underlies the local and relative spectral analyses developed below.
Within this broader family of spectral readout functionals, the conventional one-loop effective action appears as the distinguished node-coalescence limit $q\to 1$.

In Section~\ref{sec:heatkernel}, we apply this readout coordinate to the local heat-kernel hierarchy by introducing, for an elliptic operator of order $\ell$ in $d$ dimensions, a hierarchy-dependent index $\chi_{n}(d,q;\ell)$ that measures the displacement between the readout point and the candidate Mellin position associated with the $n$th heat-kernel coefficient.
Although not a new local invariant, this index also characterises the high-energy behaviour of the contribution arising from $\zeta_{A}(q-1)$ whenever the corresponding power-law spectral asymptotics exist.
The resulting framework organises poles, local special values, generic full-spectrum values, and the $q\to 1$ limit within a common Mellin coordinate.

The usefulness of this common coordinate becomes particularly clear in relative spectral problems.
Depending on whether a leading local hierarchy survives at finite order or the entire local power-law hierarchy cancels, the readout family distinguishes systems with surviving local structure from those governed purely by global spectral rearrangements.
In Section~\ref{sec:relative}, we illustrate these two cases using a reflectionless soliton and a twisted circle.
In Section~\ref{sec:4dtheory}, we show that the same framework reproduces the standard local scale response at $q=1$, while the first $q$ derivative of that scale response determines the standard zeta-regularised one-loop effective action.
Section~\ref{sec:conclusion} discusses the scope and possible extensions of the framework.

%%%%%%%%%%%%%%%%%%%%%%%%%%%%%%%%%%%%%%%%%%%%%%%%%%%%%%%%
%%%%%%%%%%%%%%%%%%%%%%%%%%%%%%%%%%%%%%%%%%%%%%%%%%%%%%%%
\section{Finite-Difference Readout and Spectral Response}\label{sec:readout}

In this section, we formulate the finite-difference readout and its associated spectral response.
The readout parameter $q$ simultaneously specifies the Mellin evaluation point $s=q-1$ and the response weight $\lambda^{-q}$; in the free-scalar example, the density of states and the readout weight combine into a single aggregation exponent.

%%===================
\subsection{Readout and Variational Response}

For a strictly positive self-adjoint elliptic operator $A$, we first define the spectral zeta function in its region of convergence by
\begin{equation}
\zeta_{A}(s)=\int_{0}^{\infty}\lambda^{-s}\dd\nu_{A}(\lambda).
\end{equation}
For a purely discrete spectrum, this reduces to
\begin{equation}
\zeta_{A}(s)=\sum_{k}\lambda_{k}^{-s}.
\end{equation}
The zeta function is then analytically continued, where necessary, to a meromorphic function of $s$.
For non-compact problems with continuous spectra, the absolute zeta function need not exist, in which case one instead uses either the volume density considered below or the relative zeta function discussed in Section~\ref{sec:relative}.
We assume throughout that $\zeta_{A}(s)$ is regular in a neighbourhood of $s=0$; zero modes, when present, require separate treatment.
If $s=q-1$ coincides with a genuine pole, no finite readout is defined at that value of $q$.
The associated pole structure is analysed in Section~\ref{sec:heatkernel} in relation to the local heat-kernel hierarchy.

For a dimensionless operator $A$, we define the one-loop readout functional corresponding to Eq.~\eqref{eq:qdef} by
\begin{equation}\label{eq:W_q}
\mathcal{W}_{q}[A]\equiv\frac{1}{2}D_{q}[A]=\frac{1}{2}\frac{\zeta_{A}(q-1)-\zeta_{A}(0)}{1-q}.
\end{equation}
For $q\neq 1$, $\mathcal{W}_{q}[A]$ is a fixed-spectrum readout rather than the standard zeta-regularised one-loop effective action; the latter is recovered at $q=1$, where $\mathcal{W}_{1}[A]$ is defined by the node-coalescence limit.
Indeed, near $q=1$,
\begin{equation}
\mathcal{W}_{q}[A]=-\frac{1}{2}\zeta_{A}'(0)-\frac{q-1}{4}\zeta_{A}''(0)+O\bigl((q-1)^{2}\bigr),
\end{equation}
and hence
\begin{equation}
\lim_{q\to 1}\mathcal{W}_{q}[A]=-\frac{1}{2}\zeta_{A}'(0).
\end{equation}
Upon extending $q$ to complex values, the readout forms a meromorphic family satisfying
\begin{equation}
\zeta_{A}(q-1)=\zeta_{A}(0)+2(1-q)\mathcal{W}_{q}[A],\quad q\neq 1.
\end{equation}
Thus, the family provides an exact reparametrisation of the information contained in $\zeta_{A}(q-1)$ along the common $q$ coordinate with the fixed node at $s=0$.
In the spectral-sensitivity comparisons below, we restrict $q$ to real values for which $\zeta_{A}(q-1)$ is regular.

To expose the eigenvalue-level content of the readout and prepare its variational interpretation, we introduce
\begin{equation}\label{eq:lam_aggr}
F_{q}(\lambda)=\frac{1}{2}\frac{\lambda^{1-q}-1}{1-q}.
\end{equation}
Whenever the corresponding eigenvalue sum converges,
\begin{equation}
\mathcal{W}_{q}[A]=\sum_{k}F_{q}(\lambda_{k}).
\end{equation}
This identity holds exactly in finite dimensions, while in infinite dimensions the sum is understood through the analytic continuation defining Eq.~\eqref{eq:W_q}.
The aggregation function satisfies
\begin{equation}\label{eq:F}
\lim_{q\to 1}F_{q}(\lambda)=\frac{1}{2}\ln\lambda,\quad F_{q}'(\lambda)=\frac{1}{2}\lambda^{-q}.
\end{equation}
Thus, $F_{q}(\lambda)$ gives the elementary contribution of an eigenvalue to the readout, while its derivative supplies the response weight used below.
Apart from the factor $\frac{1}{2}$ and the parameter convention, $F_{q}$ has the same functional form as the $q$-logarithm of Tsallis statistics, $\ln_{q}x\equiv2F_{q}(x)$ \cite{Tsallis88}.
This is only an algebraic analogy; no probabilistic or thermodynamic interpretation is assigned to the operator spectrum.
The exact finite-dimensional correspondence with $\Tr\ln_{q}A$, together with its variational and relative forms, is presented in Appendix~\ref{sec:q-alg}.

With this eigenvalue-level structure in place, consider a smooth family of strictly positive self-adjoint elliptic operators $A(\tau)$ with a common domain, remaining strictly positive throughout the variation.
We assume that $\delta\zeta_{A}(s)$ extends meromorphically to neighbourhoods of $s=0$ and $s=q-1$, and that the variation may be interchanged with this continuation.
In the region where the ordinary operator trace converges, one has
\begin{equation}
\delta\zeta_{A}(s)=-s\Tr\bigl(A^{-s-1}\delta A\bigr).
\end{equation}
We denote the meromorphic continuation of this trace expression by
\begin{equation}
\mathcal{T}_{A}(s;\delta A)\equiv-\frac{1}{s}\delta\zeta_{A}(s),\quad s\neq 0,
\end{equation}
using $\mathcal{T}_{A}$ only as shorthand and not as an independently defined regularised trace.
For $q\neq 1$, the complete variation of Eq.~\eqref{eq:W_q} is
\begin{equation}\label{eq:W_q_var}
\delta\mathcal{W}_{q}[A]=\frac{1}{2}\frac{\delta\zeta_{A}(q-1)-\delta\zeta_{A}(0)}{1-q}=\frac{1}{2}\mathcal{T}_{A}(q-1;\delta A)-\frac{1}{2(1-q)}\delta\zeta_{A}(0).
\end{equation}
The first term represents the response arising from $\zeta_{A}(q-1)$ and, for a discrete spectrum in the region of ordinary trace convergence, may be written as
\begin{equation}
\frac{1}{2}\Tr\bigl(A^{-q}\delta A\bigr)=\frac{1}{2}\sum_{k}\lambda_{k}^{-q}\delta\lambda_{k}.
\end{equation}
Comparison with Eq.~\eqref{eq:F} shows that the per-eigenvalue response is $F_{q}'(\lambda)\delta\lambda=\frac{1}{2}\lambda^{-q}\delta\lambda$.
Thus, apart from the conventional one-loop factor $\frac{1}{2}$, the spectral response weight is $\lambda^{-q}$.
The same parameter $q$ therefore controls both the aggregation of eigenvalues in the readout and their weighting in the variational response.
The complete variation also contains the reference contribution from $\delta\zeta_{A}(0)$, so the eigenvalue-sum expression represents the full variation only when $\delta\zeta_{A}(0)=0$.
Otherwise, it represents the $q$-dependent component separately.

Furthermore, as $q\to 1$,
\begin{equation}
\mathcal{T}_{A}(q-1;\delta A)=-\frac{\delta\zeta_{A}(0)}{q-1}-\delta\zeta_{A}'(0)+O(q-1).
\end{equation}
Accordingly, the two singular terms appearing in Eq.~\eqref{eq:W_q_var} cancel each other, and
\begin{equation}
\lim_{q\to 1}\delta\mathcal{W}_{q}[A]=-\frac{1}{2}\delta\zeta_{A}'(0),
\end{equation}
thereby recovering the variation of the standard zeta-regularised one-loop effective action.
As $q$ is varied away from $q=1$, the spectral weight in the $q$-dependent component of the variation changes continuously.
At fixed reference scale and nondimensionalisation, increasing $q$ suppresses the response weight assigned to higher eigenvalues relative to $\lambda=1$, while for positive $q$ the response to eigenvalues approaching zero is enhanced.
Thus, varying $q$ continuously redistributes the spectral sensitivity under a fixed scaling convention.

The dependence on the reference scale introduced by the nondimensionalisation of $A$ can be examined separately.
For a dimensionful operator $\widetilde{A}$ of order $\ell$, let $A=\widetilde{A}/\mu^{\ell}$.
Taking the explicit scale derivative at fixed $\widetilde{A}$ gives
\begin{equation}\label{eq:gen_scale_resp}
\mu\frac{\partial\mathcal{W}_{q}[A]}{\partial\mu}=-\frac{\ell}{2}\zeta_{A}(q-1).
\end{equation}
Whereas the scale response at $q=1$ is determined by the local special value $\zeta_{A}(0)$, at generic regular values of $q$ it depends on $\zeta_{A}(q-1)$ and hence generally on the full spectrum.
The following free-scalar example makes this distinction explicit.

%%===================
\subsection{Free Scalar Response and Spectral Aggregation}\label{sec:scalar}

The spectral response weight $\lambda^{-q}$ admits a natural real-space representation.
For the massless Laplace operator $-\Delta$ on a $d$-dimensional flat space, under the standard Fourier transform convention, one has, for $0<\Re q<d/2$,
\begin{equation}
\langle x|(-\Delta)^{-q}|y\rangle=\frac{\Gamma\bigl(\frac{d}{2}-q\bigr)}{4^{q}\pi^{d/2}\Gamma(q)}|x-y|^{2q-d},\quad x\neq y.
\end{equation}
Thus, the readout coordinate $q$ determines both the spectral-space weight $\lambda^{-q}$ and the real-space distance exponent $2q-d$.
At $q=d/2$, the distance exponent vanishes while the coefficient $\Gamma\bigl(\frac{d}{2}-q\bigr)$ has a pole.
For the dimensionless operator $(-\Delta)/\mu^{2}$, the position-dependent term in the finite part of the Laurent expansion is
\begin{equation}
-\frac{2\mu^{d}}{4^{d/2}\pi^{d/2}\Gamma\bigl(\frac{d}{2}\bigr)}\ln\!\left(\mu|x-y|\right),\quad x\neq y.
\end{equation}
Accordingly, $q=d/2$ marks the transition of the real-space response from power-law to logarithmic behaviour.
The additive constant depends on the treatment of the zero mode and on the infrared convention, while $\mu$ is the fixed reference scale.

We now consider the massive operator
\begin{equation}
\widetilde{A}=-\Delta+m^{2},\quad m>0,\quad A=\frac{\widetilde{A}}{\mu^{2}}.
\end{equation}
Its real-space kernel is exponentially suppressed as $e^{-m|x-y|}$ at large distances, with decay length $m^{-1}$.
The parameter $q$ therefore does not set the large-distance decay scale, but controls the short-distance behaviour, the algebraic prefactor, and the relative weighting of spectral regions.
Using box normalisation as an infrared regulator and then taking the infinite-volume limit, the spectral-zeta density is, for $\Re s>d/2$,
\begin{equation}
\frac{\zeta_{A}(s)}{V}=\int\frac{\dd^{d}p}{(2\pi)^{d}}\left(\frac{p^{2}+m^{2}}{\mu^{2}}\right)^{-s}=\frac{\mu^{2s}}{(4\pi)^{d/2}}\frac{\Gamma\bigl(s-\frac{d}{2}\bigr)}{\Gamma(s)}(m^{2})^{d/2-s},
\end{equation}
and the right-hand side gives its meromorphic continuation.
The power-law component of the finite-difference readout arising from $\zeta_{A}(q-1)$ is
\begin{equation}
\frac{1}{2(1-q)}\frac{\zeta_{A}(q-1)}{V}=\frac{\mu^{d}}{2(1-q)(4\pi)^{d/2}}\frac{\Gamma\bigl(q-1-\frac{d}{2}\bigr)}{\Gamma(q-1)}\left(\frac{m^{2}}{\mu^{2}}\right)^{\nu(d,q)},
\end{equation}
where
\begin{equation}\label{eq:nu}
\nu(d,q)=\frac{d}{2}-q+1.
\end{equation}
The exponent $\nu(d,q)$ is the mass power produced by combining the leading Weyl density of states with the readout weight $\lambda^{1-q}$.
It is a spectral aggregation exponent for this free-field example rather than a critical exponent of the underlying theory.
For fixed $q$ with finite nonzero prefactor, the power-law component vanishes in the massless limit when $\nu(d,q)>0$ and becomes infrared singular when $\nu(d,q)<0$.
The condition $\nu(d,q)=0$ occurs at $q=d/2+1$ and marks the transition between power-law and logarithmic behaviour for the leading Weyl contribution to the readout.
This lies one unit above the real-space response threshold at $q=d/2$, reflecting the difference between the readout weight $\lambda^{1-q}$ and the response weight $\lambda^{-q}$.
At $q=d/2+1$, the evaluation point $s=q-1=d/2$ is a genuine pole of the zeta function, so $\mathcal{W}_{q}[A]$ is not finite there, while the finite part of its expansion contains logarithmic dependence on $m^{2}/\mu^{2}$.
The next section shows that this leading Weyl transition is the first example of the general heat-kernel hierarchy governing the readout.

%%%%%%%%%%%%%%%%%%%%%%%%%%%%%%%%%%%%%%%%%%%%%%%%%%%%%%%%%%%%%%%
%%%%%%%%%%%%%%%%%%%%%%%%%%%%%%%%%%%%%%%%%%%%%%%%%%%%%%%%%%%%%%%
\section{Readout Across the Heat-Kernel Hierarchy}\label{sec:heatkernel}

In this section, we place the finite-difference readout within the local heat-kernel hierarchy.
The parameter $q$ changes neither the operator nor the heat-kernel coefficients, but only the Mellin evaluation point on the fixed spectral zeta function.
This leads to a hierarchy-dependent aggregation exponent measuring the displacement between the readout point and each candidate Mellin position.
We then illustrate the resulting structure in four dimensions and under spectral power transformations.

%%===================
\subsection{Heat-Kernel--Zeta Correspondence}

We now specialise to a strictly positive self-adjoint elliptic differential operator $A$ of positive even order $\ell$ on a compact $d$-dimensional Riemannian manifold without boundary.
The short-time heat-trace expansion is given by \cite{Gilkey95,Vassilevich03,Kirsten01}
\begin{equation}
K_{A}(t)=\Tr\bigl(e^{-tA}\bigr)~\sim~\sum_{n\geq 0}a_{n}(A)t^{(n-d)/\ell},\quad t\to 0^{+}.
\end{equation}
The corresponding spectral zeta function is represented, for sufficiently large $\Re s$, by
\begin{equation}
\zeta_{A}(s)=\frac{1}{\Gamma(s)}\int_{0}^{\infty}t^{s-1}K_{A}(t)\dd t.
\end{equation}
Accordingly, the Mellin position associated with the $n$th heat-kernel term is
\begin{equation}\label{eq:heat_candidate}
s_{n}=\frac{d-n}{\ell}.
\end{equation}
When $s_{n}\notin\mathbb{Z}_{\leq 0}$ and $a_{n}(A)\neq 0$, this position is a genuine simple pole of $\zeta_{A}(s)$,
\begin{equation}
\Res_{s=s_{n}}\zeta_{A}(s)=\frac{a_{n}(A)}{\Gamma(s_{n})}.
\end{equation}
By contrast, when $s_{n}=-k$ with $k\in\mathbb{Z}_{\geq 0}$, the pole is cancelled by the zero of $\Gamma(s)^{-1}$, and the same local coefficient instead appears as the regular special value
\begin{equation}
\zeta_{A}(-k)=(-1)^{k}k!\,a_{d+k\ell}(A).
\end{equation}
At generic regular points, on the other hand, $\zeta_{A}(s)$ cannot, in general, be reduced to a finite set of local coefficients but depends on the full spectrum, including its low-energy part.

The finite-difference readout identifies the evaluation point as $s=q-1$, so the $n$th heat-kernel term is encountered at the candidate readout position $q=1+s_{n}=1+(d-n)/\ell$.
Varying $q$ moves continuously along the Mellin axis and passes through the positions associated with successive local heat-kernel hierarchies.
The node-coalescence point $q=1$ provides the simplest illustration.
At the corresponding Mellin position $s=0$, the regular value $\zeta_{A}(0)=a_{d}(A)$ is fixed by the local heat-kernel coefficient, whereas the node-coalescence limit gives $\mathcal{W}_{1}[A]=-\frac{1}{2}\zeta_{A}'(0)$, which generally depends on the full spectrum.
The readout family thus places both local heat-kernel data and the standard zeta-regularised one-loop contribution within a common Mellin hierarchy.

%%===================
\subsection{Hierarchy-Dependent Aggregation Exponents}

We extend the leading aggregation exponent $\nu(d,q)$ obtained for the free scalar field in the preceding section to the general heat-kernel hierarchy.
We define the difference between the candidate Mellin position Eq.~\eqref{eq:heat_candidate} of the $n$th hierarchy and the readout point $s=q-1$ by
\begin{equation}
\chi_{n}(d,q;\ell)\equiv s_{n}-(q-1)=\frac{d-n}{\ell}-q+1.
\end{equation}
The quantity $\chi_{n}$ measures the displacement of the readout point from the candidate Mellin position of the $n$th local hierarchy.
At the level of Mellin positions, the same displacement is formally equivalent to the shift $d\mapsto d-\ell(q-1)$, suggesting an effective-dimensional interpretation of the readout parameter without implying any actual change in spacetime dimension.
Its further interpretation as a high-energy aggregation exponent requires that $s_{n}\notin\mathbb{Z}_{\leq 0}$ and that the smoothed spectral density contains an asymptotic component
\begin{equation}
\rho_{n}(\lambda)\sim C_{n}\lambda^{s_{n}-1},\quad C_{n}=\frac{a_{n}(A)}{\Gamma(s_{n})}.
\end{equation}
Here, $\rho_{n}(\lambda)$ denotes the $n$th component of the smoothed high-energy asymptotics.
The coefficient relation follows by matching the Laplace transform of the asymptotic density term to the heat-kernel contribution $a_{n}(A)t^{-s_{n}}$.
Under this assumption, the corresponding spectral-density integrand in $\zeta_{A}(q-1)$ behaves as
\begin{equation}
\lambda^{1-q}\rho_{n}(\lambda)~\sim~C_{n}\lambda^{\chi_{n}-1}.
\end{equation}
Restricting the high-energy region to $\lambda_{0}\leq\lambda\leq\lambda_{\max}$ gives
\begin{equation}
\int_{\lambda_{0}}^{\lambda_{\max}}\lambda^{1-q}\rho_{n}(\lambda)\dd\lambda~\sim~
\begin{cases}
~\displaystyle\frac{C_{n}}{\chi_{n}}
\bigl(\lambda_{\max}^{\chi_{n}}-\lambda_{0}^{\chi_{n}}\bigr)
& (\chi_{n}\neq 0),\\[1em]
~\displaystyle C_{n}\ln\frac{\lambda_{\max}}{\lambda_{0}}
& (\chi_{n}=0).
\end{cases}
\end{equation}
Thus, $\chi_{n}>0$, $\chi_{n}=0$, and $\chi_{n}<0$ correspond respectively to power-law growth, logarithmic growth, and a finite high-energy limit with power-law corrections for the contribution arising from $\zeta_{A}(q-1)$.
The condition $\chi_{n}=0$ identifies only the coincidence of the readout point with $s_{n}$.
Whether this coincidence gives a genuine pole, a regular special value when $s_{n}\in\mathbb{Z}_{\leq 0}$, or no contribution when $a_{n}(A)=0$ must be determined from the analytic structure of $\zeta_{A}(s)$.

For a Laplace-type operator, $\ell=2$, so that
\begin{equation}
\chi_{n}(d,q;2)=\frac{d-n}{2}-q+1=\nu(d,q)-\frac{n}{2}.
\end{equation}
Thus, the free-field exponent $\nu(d,q)$ introduced in Eq.~\eqref{eq:nu} is precisely the leading-hierarchy exponent $\chi_{0}(d,q;2)$, while successive local hierarchies are shifted from it by $n/2$.

The variational response carries an additional high-energy factor determined by the eigenvalue variation.
Writing $\lambda$ for the high-energy spectral variable, assume that
\begin{equation}
\delta\lambda=O\bigl(\lambda^{r/\ell}\bigr),\quad \lambda\to\infty,
\end{equation}
where $r$ characterises the effective high-energy order of the perturbation.
This scaling assumption depends on the operator family and does not follow solely from assigning differential order $r$ to $\delta A$.
Combining it with the smoothed spectral density gives
\begin{equation}
\lambda^{-q}\rho_{n}(\lambda)\delta\lambda=O\bigl(\lambda^{(d-n+r)/\ell-q-1}\bigr),
\end{equation}
and the corresponding aggregation exponent is
\begin{equation}
\chi_{n}^{(r)}(d,q;\ell)=\frac{d-n+r}{\ell}-q=\chi_{n}(d,q;\ell)+\frac{r}{\ell}-1.
\end{equation}
Thus, taking a variation shifts the hierarchy-dependent aggregation exponent by $r/\ell-1$, so perturbations of different effective order may exhibit different high-energy behaviour at the same readout position.

%%===================
\subsection{Four-Dimensional Laplace-Type Operators}

We now specialise the preceding arrangement of candidate evaluation points to the four-dimensional theories considered below.
For a Laplace-type operator with $d=4$ and $\ell=2$, the candidate readout position associated with the $n$th heat-kernel term is $q=3-n/2$.
On a manifold without boundary, when the odd-numbered heat-kernel coefficients vanish, $n=0,\,2,\,4,\,6,\ldots$ correspond to $q=3,\,2,\,1,\,0,\ldots$.
At $q=3$, a genuine pole occurs at $s=2$, determined by $a_{0}(A)$, while at $q=2$ a genuine pole occurs at $s=1$ when $a_{2}(A)\neq 0$.
At $q=1$, the regular special value is $\zeta_{A}(0)=a_{4}(A)$, while the node-coalescence limit of $\mathcal{W}_{q}[A]$ yields the standard zeta-regularised one-loop contribution $-\frac{1}{2}\zeta_{A}'(0)$.
At $q=0,\,-1,\,-2,\ldots$, higher-order local coefficients appear as regular special values, whereas generic intermediate values depend on the full spectrum.
This four-dimensional arrangement is summarised in Table~\ref{tab:qhierarchy}.

%++++++++++++++++++++++++++++++++++++++++++++++++++++
\begin{table}[t]
\centering
\caption{Correspondence between the readout parameter $q$, the Mellin evaluation point $s=q-1$, and the local heat-kernel hierarchy for a second-order elliptic operator in $d=4$.}
\vspace{0.3em}
\label{tab:qhierarchy}
{\footnotesize
\setlength{\tabcolsep}{8pt}
\begin{tabular}{cccll}
\toprule
Local coefficient & $q$ & $s=q-1$ & $\zeta_{A}(s)$ & Spectral meaning at $s=q-1$\\
\midrule
$a_{0}$ & $3$ & $2$ & Genuine pole & Volume divergence\\
$a_{2}$ & $2$ & $1$ & Genuine pole if non-vanishing & Dimension-two local term\\
$a_{4}$ & $1$ & $0$ & Regular special value & $\zeta_{A}(0)=a_{4}(A)$; local scale response\\
$a_{6}$ & $0$ & $-1$ & Regular special value & $\zeta_{A}(-1)=-a_{6}(A)$\\
\bottomrule
\end{tabular}
}
\end{table}
%++++++++++++++++++++++++++++++++++++++++++++++++++++

%%===================
\subsection{Spectral-Power Covariance}

The Mellin positions and aggregation exponents introduced above transform naturally under spectral power maps.
We therefore consider the transformation $A\mapsto A^{\theta}$ with $\theta>0$, defined by spectral functional calculus.
Since the eigenvalues of $A^{\theta}$ are $\{\lambda_{k}^{\,\theta}\}$,
\begin{equation}\label{eq:power_zeta_cov}
\zeta_{A^{\theta}}(s)=\zeta_{A}(\theta s).
\end{equation}
Defining
\begin{equation}
q'=1+\theta(q-1),
\end{equation}
one obtains
\begin{equation}\label{eq:power_W_cov}
\mathcal{W}_{q}[A^{\theta}]=\theta\mathcal{W}_{q'}[A].
\end{equation}
Thus, the spectral mapping $\lambda\mapsto\lambda^{\theta}$ rescales the Mellin evaluation point according to $q-1\mapsto q'-1=\theta(q-1)$ and maps the readout at $q$ for $A^{\theta}$ to that at $q'$ for $A$.
The point $q=1$ is a fixed point of this transformation, and the node-coalescence limit of Eq.~\eqref{eq:power_W_cov} gives $\mathcal{W}_{1}[A^{\theta}]=\theta\mathcal{W}_{1}[A]$.

It follows from Eq.~\eqref{eq:power_zeta_cov} that the poles of the zeta function obey the same scaling law.
If $\zeta_{A}(s)$ has a simple pole at $s=\sigma$, the corresponding pole of $\zeta_{A^{\theta}}(s)$ is located at $s=\sigma/\theta$, and its residue transforms as
\begin{equation}
\Res_{s=\sigma/\theta}\zeta_{A^{\theta}}(s)=\frac{1}{\theta}\Res_{u=\sigma}\zeta_{A}(u).
\end{equation}
Accordingly, the readout point and the pole set move under the same rescaling.

The same rescaling extends to the hierarchy-dependent aggregation exponents.
If $A$ has order $\ell$, then $A^{\theta}$ has order $\ell\theta$, and
\begin{equation}
\chi_{n}(d,q;\ell\theta)=\frac{1}{\theta}\chi_{n}(d,q';\ell).
\end{equation}
This identity describes the algebraic rescaling of the Mellin positions and readout coordinate.
For non-integer $\theta$, however, $A^{\theta}$ is generally an elliptic pseudodifferential operator rather than a differential operator \cite{Seeley67}, so the relation should not be interpreted as preserving the local differential hierarchy or physical equivalence.

%%%%%%%%%%%%%%%%%%%%%%%%%%%%%%%%%%%%%%%%%%%%%%%%%%%%%%%%%%%%%%%
%%%%%%%%%%%%%%%%%%%%%%%%%%%%%%%%%%%%%%%%%%%%%%%%%%%%%%%%%%%%%%%
\section{Relative Readout and Local-Hierarchy Cancellation}\label{sec:relative}

The preceding section organised the readout relative to the local heat-kernel hierarchy of a single operator.
We now consider differences between two operators, where common local contributions may cancel before the remaining spectrum is read out.
The relative readout therefore distinguishes problems with a surviving leading local hierarchy from those in which the entire local power-law hierarchy disappears.

%%===================
\subsection{Relative Readout and Surviving Hierarchies}

Consider a pair of strictly positive self-adjoint operators $(A,A_{0})$ defined on the same space, and suppose that the relative heat trace
\begin{equation}
\Delta K_{A;A_{0}}(t)=\Tr\bigl(e^{-tA}-e^{-tA_{0}}\bigr)
\end{equation}
is well defined.
We define the corresponding relative zeta function, initially in its region of convergence, by
\begin{equation}
\Delta\zeta_{A;A_{0}}(s)=\frac{1}{\Gamma(s)}\int_{0}^{\infty}t^{s-1}\Delta K_{A;A_{0}}(t)\dd t,
\end{equation}
and then analytically continue it in $s$.
The relative finite-difference readout is
\begin{equation}\label{eq:rel_q_zeta}
\Delta\mathcal{W}_{q}[A;A_{0}]=\frac{1}{2}\frac{\Delta\zeta_{A;A_{0}}(q-1)-\Delta\zeta_{A;A_{0}}(0)}{1-q}.
\end{equation}
Whenever the individual zeta functions exist,
\begin{equation}
\Delta\zeta_{A;A_{0}}(s)=\zeta_{A}(s)-\zeta_{A_{0}}(s),
\end{equation}
and the corresponding readout satisfies
\begin{equation}
\Delta\mathcal{W}_{q}[A;A_{0}]=\mathcal{W}_{q}[A]-\mathcal{W}_{q}[A_{0}].
\end{equation}
For non-compact or scattering problems, however, the individual zeta functions may fail to exist even when the relative quantities remain well defined \cite{Muller98,Yafaev92}.

Suppose that the operator pair satisfies the assumptions required for Krein's trace formula and that the spectral shift function $\xi_{A;A_{0}}(\lambda)$ is defined \cite{Yafaev92}.
We fix its sign by
\begin{equation}
\Tr\bigl[f(A)-f(A_{0})\bigr]=\int_{0}^{\infty}\xi_{A;A_{0}}(\lambda)\dd f(\lambda).
\end{equation}
With this convention, choosing $f(\lambda)=e^{-t\lambda}$ gives
\begin{equation}
\Delta K_{A;A_{0}}(t)=-t\int_{0}^{\infty}e^{-t\lambda}\xi_{A;A_{0}}(\lambda)\dd\lambda.
\end{equation}
Taking the Mellin transform and interchanging the integrations in the region of direct convergence then yields
\begin{equation}\label{eq:rel_zeta_spect_shift}
\Delta\zeta_{A;A_{0}}(s)=-s\int_{0}^{\infty}\lambda^{-s-1}\xi_{A;A_{0}}(\lambda)\dd\lambda.
\end{equation}
Thus, the heat-trace and spectral-shift representations provide two equivalent forms of the same relative zeta function.
For discrete spectra, let $N_{A}(\lambda)$ denote the number of eigenvalues of $A$ not exceeding $\lambda$, counted with multiplicity.
The sign convention above then gives $\xi_{A;A_{0}}(\lambda)=N_{A_{0}}(\lambda)-N_{A}(\lambda)$.
In scattering problems, the spectral shift function instead combines bound-state jumps with the continuous shift determined by the scattering phase.
Substitution into Eq.~\eqref{eq:rel_q_zeta} gives
\begin{equation}\label{eq:spect_shift_q_gen}
\Delta\mathcal{W}_{q}[A;A_{0}]=\frac{1}{2}\int_{0}^{\infty}\lambda^{-q}\xi_{A;A_{0}}(\lambda)\dd\lambda-\frac{\Delta\zeta_{A;A_{0}}(0)}{2(1-q)}.
\end{equation}
When $\Delta\zeta_{A;A_{0}}(0)=0$, the relative readout is given entirely by the Mellin-weighted spectral shift.
When the ordinary integral does not converge, Eq.~\eqref{eq:rel_q_zeta} defines the relative readout, and Eq.~\eqref{eq:spect_shift_q_gen} is understood by analytic continuation from the region of direct convergence.
The standard relative determinant compresses the spectral rearrangement into a single logarithmic quantity, whereas varying $q$ redistributes the sensitivity among bound states, threshold behaviour, and high-energy spectral components.

The high-energy structure of this relative readout is determined by the short-time behaviour of the relative heat trace.
Suppose that $\Delta K_{A;A_{0}}(t)$ admits a short-time expansion in powers $t^{(n-d)/\ell}$ with relative coefficients $\Delta a_{n}$; whenever the individual coefficients are defined, $\Delta a_{n}=a_{n}(A)-a_{n}(A_{0})$.
When the two operators have the same order and principal symbol, the leading Weyl term typically cancels, while the surviving higher coefficients reflect differences in lower-order symbols, potentials, connections, curvatures, or boundary data.
Let $n_{\ast}$ be the first index for which $\Delta a_{n_{\ast}}\neq 0$.
The leading short-time behaviour is then
\begin{equation}
\Delta K_{A;A_{0}}(t)\sim\Delta a_{n_{\ast}}t^{(n_{\ast}-d)/\ell},\quad t\to 0^{+},
\end{equation}
and the corresponding Mellin position is
\begin{equation}
s_{\ast}=\frac{d-n_{\ast}}{\ell}.
\end{equation}
The leading relative aggregation exponent is therefore
\begin{equation}\label{eq:rel_aggr}
\chi_{n_{\ast}}(d,q;\ell)=s_{\ast}-(q-1)=\frac{d-n_{\ast}}{\ell}-q+1.
\end{equation}
Under the smoothed spectral-asymptotic assumptions used in Section~\ref{sec:heatkernel}, this exponent classifies the high-energy behaviour of the leading power-law contribution arising from $\Delta\zeta_{A;A_{0}}(q-1)$.
Cancellation of the lower relative coefficients therefore shifts the leading high-energy sensitivity to the first surviving hierarchy.
If all relative heat-kernel coefficients vanish and
\begin{equation}
\Delta K_{A;A_{0}}(t)=O\bigl(t^{N}\bigr),\quad t\to 0^{+},
\end{equation}
for every $N>0$, no finite $n_{\ast}$ exists.
The relative zeta function then has no pole sequence generated by local short-time power terms, and the readout probes spectral differences beyond the entire local power-law hierarchy.

%%===================
\subsection{Two Relative Models}\label{subsec:rel_models}

We illustrate the two cases identified above with a reflectionless soliton and a twisted circle.
For the soliton, a first surviving relative hierarchy produces a genuine pole at a distinguished readout position, whereas for the twisted circle all local power-law coefficients cancel, leaving only global winding-sector contributions.
For both examples, the heat traces are written for dimensionful operator pairs; nondimensionalisation by $\mu$ rescales the nonzero coefficients but leaves their vanishing pattern, hierarchy indices, and associated Mellin positions unchanged.
Detailed derivations of the scattering integrals, Poisson resummation, and analytic continuations are given in Appendix~\ref{sec:relative_examples_details}.

%---------
\subsubsection{Reflectionless Soliton: Survival of a Relative Hierarchy}

Consider the operator pair on the real line
\begin{equation}\label{eq:soliton}
\widetilde{A}_{\mathrm{sol}}=-\frac{\dd^{2}}{\dd x^{2}}+M^{2}-2\kappa^{2}\sech^{2}(\kappa x),\quad\widetilde{A}_{0}=-\frac{\dd^{2}}{\dd x^{2}}+M^{2},\quad M>\kappa>0.
\end{equation}
The operator $\widetilde{A}_{\mathrm{sol}}$ has one bound state with eigenvalue $M^{2}-\kappa^{2}$ and a continuous spectrum characterised by the scattering phase $\delta(k)=2\arctan(\kappa/k)$, whose derivative gives the continuum density shift
\begin{equation}
\frac{1}{\pi}\frac{\dd\delta(k)}{\dd k}=-\frac{2\kappa}{\pi(k^{2}+\kappa^{2})}.
\end{equation}
Under the sign convention of Eq.~\eqref{eq:rel_zeta_spect_shift}, the downward unit jump at the bound-state eigenvalue and the continuous contribution determined by the scattering phase give
\begin{equation}\label{eq:sol_rel_zeta_start}
\Delta\zeta_{\mathrm{sol}}(s)=\left(\frac{M^{2}-\kappa^{2}}{\mu^{2}}\right)^{-s}+\frac{1}{\pi}\int_{0}^{\infty}\left(\frac{k^{2}+M^{2}}{\mu^{2}}\right)^{-s}\frac{\dd\delta(k)}{\dd k}\dd k.
\end{equation}
The explicit piecewise form of the spectral shift function is given in Eq.~\eqref{eq:sol_spectral_shift} of Appendix~\ref{sec:relative_examples_details}, where Eq.~\eqref{eq:sol_rel_zeta_start} is also derived.
Using the same decomposition, one obtains the exact relative heat trace of the dimensionful pair, whose short-time behaviour is
\begin{equation}\label{eq:rel_ker_sol}
\Delta K_{\mathrm{sol}}(t)\equiv\Tr\bigl(e^{-t\widetilde{A}_{\mathrm{sol}}}-e^{-t\widetilde{A}_{0}}\bigr)=e^{-(M^{2}-\kappa^{2})t}\erf\bigl(\kappa\sqrt{t}\bigr)~\sim~\frac{2\kappa}{\sqrt{\pi}}\,t^{1/2}+O\bigl(t^{3/2}\bigr),\quad t\to 0^{+}.
\end{equation}
Thus, for the heat trace of the dimensionful pair in Eq.~\eqref{eq:rel_ker_sol}, the first nonzero relative coefficient occurs at $n_{\ast}=2$, with $\Delta\widetilde{a}_{2}=2\kappa/\sqrt{\pi}$.
Because nondimensionalisation leaves the hierarchy index unchanged, substituting $n_{\ast}=2$, $d=1$, and $\ell=2$ into Eq.~\eqref{eq:rel_aggr} gives
\begin{equation}
\chi_{\mathrm{sol}}(q)=\frac{1}{2}-q.
\end{equation}
The first surviving relative hierarchy therefore selects the distinguished value $q=1/2$.
For $q>1/2$, the leading high-energy contribution arising from $\Delta\zeta_{\mathrm{sol}}(q-1)$ is convergent.
At $q=1/2$, the readout reaches the pole associated with the first surviving hierarchy, so no finite $\Delta\mathcal{W}_{q}$ is defined there.
The scattering calculation in Appendix~\ref{sec:relative_examples_details} gives $\Delta\zeta_{\mathrm{sol}}(0)=0$ and the meromorphic continuation required to evaluate the readout.
The resulting relative finite-difference readout is
\begin{equation}\label{eq:sol_rel_q}
\Delta\mathcal{W}_{q}=\frac{1}{2(1-q)}\Biggl[\left(\frac{M^{2}-\kappa^{2}}{\mu^{2}}\right)^{1-q}-\frac{\Gamma\bigl(q-\frac{1}{2}\bigr)}{\sqrt{\pi}\,\Gamma(q)}\left(\frac{M^{2}}{\mu^{2}}\right)^{1-q}{}_{2}\mathrm{F}_{1}\!\left(q-1,\frac{1}{2};q;1-\frac{\kappa^{2}}{M^{2}}\right)\Biggr].
\end{equation}
This expression is obtained first for $q>1/2$ and then continued meromorphically in $q$.

A second distinguished value is the node-coalescence point $q=1$, where the finite-difference family reduces to one-half of the logarithm of the relative zeta determinant:
\begin{equation}\label{eq:sol_rel_det}
\Delta\mathcal{W}_{1}=\frac{1}{2}\ln\frac{M-\kappa}{M+\kappa}.
\end{equation}
This is consistent with standard scattering and Gel'fand--Yaglom evaluations of one-dimensional functional determinants \cite{Dunne08}.
The two distinguished values have different origins: $q=1/2$ is the local UV pole fixed by the first surviving relative hierarchy, whereas $q=1$ yields the finite full-spectrum relative one-loop contribution combining the bound state and continuum rearrangement.

The general-$q$ family also makes explicit the approach of the bound state to zero.
As $M\to\kappa^{+}$, the dimensionless bound-state eigenvalue $\bigl(M^{2}-\kappa^{2}\bigr)/\mu^{2}$ vanishes.
The factor associated with its power-law part diverges for $q>1$, becomes logarithmically singular at $q=1$, and tends to zero for $q<1$.
The complete eigenvalue contribution remains finite for $q<1$, while the complete relative zero-mode limit must be evaluated from Eq.~\eqref{eq:sol_rel_q} because the continuum contribution remains present.
Thus, varying $q$ generates neither the local pole nor the relative determinant, but organises the local-hierarchy pole, the standard relative one-loop limit, and the infrared approach to a zero mode within a single meromorphic coordinate.

%---------
\subsubsection{Twisted Circle: Cancellation of the Entire Local Hierarchy}

For the complementary example, let $\alpha\in\mathbb{R}/\mathbb{Z}$ be a twist parameter on a circle of length $L$, and consider the operator
\begin{equation}
\widetilde{A}_{\alpha}=-\partial_{x}^{2}+m^{2},\quad\phi(L)=e^{2\pi i\alpha}\phi(0),\quad\phi'(L)=e^{2\pi i\alpha}\phi'(0).
\end{equation}
The corresponding eigenvalues are
\begin{equation}
\lambda_{n,\alpha}=\left(\frac{2\pi(n+\alpha)}{L}\right)^{2}+m^{2},\quad n\in\mathbb{Z}.
\end{equation}
We take $m>0$ and compare $\widetilde{A}_{\alpha}$ with the periodic reference operator $\widetilde{A}_{0}$ obtained by setting $\alpha=0$.
The pair has the same local differential expression and differs only through the global holonomy.
Defining $A_{\alpha}=\widetilde{A}_{\alpha}/\mu^{2}$ and $A_{0}=\widetilde{A}_{0}/\mu^{2}$, and setting $c=mL/(2\pi)$, the relative zeta function is initially represented by
\begin{equation}\label{eq:twist_rel_zeta_start}
\Delta\zeta_{\alpha}(s)=\left(\frac{\mu L}{2\pi}\right)^{2s}\sum_{n\in\mathbb{Z}}\left\{\left[(n+\alpha)^{2}+c^{2}\right]^{-s}-\left[n^{2}+c^{2}\right]^{-s}\right\}.
\end{equation}
Applying Poisson resummation to the relative heat trace
$\Delta K_{\alpha}(t)\equiv\Tr\bigl(e^{-t\widetilde{A}_{\alpha}}-e^{-t\widetilde{A}_{0}}\bigr)$ gives the winding-number expansion
\begin{equation}\label{eq:twist_rel_heat_compact}
\Delta K_{\alpha}(t)=\frac{L}{\sqrt{\pi t}}e^{-m^{2}t}\sum_{k=1}^{\infty}\left[\cos(2\pi k\alpha)-1\right]e^{-L^{2}k^{2}/(4t)}.
\end{equation}
For every $0<c_{0}<L^{2}/4$,
\begin{equation}\label{eq:rel_ker_twist_zero}
\Delta K_{\alpha}(t)=O\bigl(e^{-c_{0}/t}\bigr), \quad t\to 0^{+}.
\end{equation}
Thus, every coefficient in the ordinary local heat-kernel expansion vanishes, so no first surviving local coefficient exists at any finite order and no hierarchy-dependent pole arises.
The relative spectrum is instead encoded entirely by the nonzero winding sectors.
Appendix~\ref{sec:relative_examples_details} shows that $\Delta\zeta_{\alpha}(s)$ extends to an entire function of $s$ and satisfies $\Delta\zeta_{\alpha}(0)=0$.
Consequently, the relative finite-difference readout is
\begin{equation}\label{eq:twist_rel_q}
\Delta\mathcal{W}_{q}(\alpha)=\frac{2\sqrt{\pi}}{1-q}\left(\frac{\mu L}{2\pi}\right)^{2(q-1)}\frac{1}{\Gamma(q-1)}\sum_{k=1}^{\infty}\left[\cos(2\pi k\alpha)-1\right]\left(\frac{2\pi^{2}k}{mL}\right)^{q-\frac{3}{2}}\mathrm{K}_{q-\frac{3}{2}}(kmL).
\end{equation}
Here, $\mathrm{K}_{\nu}(z)$ denotes the modified Bessel function of the second kind.
The apparent singularity at $q=1$ is removable, and $\Delta\mathcal{W}_{q}(\alpha)$ extends to an entire function of $q$.
Its value at $q=1$ is the purely global relative one-loop contribution
\begin{equation}\label{eq:twist_rel_det}
\Delta\mathcal{W}_{1}(\alpha)=\frac{1}{2}\ln\left[\frac{\cosh(mL)-\cos(2\pi\alpha)}{\cosh(mL)-1}\right].
\end{equation}
The determinant ratio inside the logarithm is purely global, since it is generated solely by changing the holonomy while leaving the local differential expression unchanged.
Away from $q=1$, varying $q$ only redistributes the nonzero winding-sector contributions through the coupled dependence of the winding power and the order of the modified Bessel function.
The readout is invariant under $\alpha\mapsto\alpha+1$ and $\alpha\mapsto 1-\alpha$, and vanishes at $\alpha=0$.

The contrast between the two models is therefore analytic as well as spectral: the reflectionless soliton has a pole at $q=1/2$ generated by a surviving local hierarchy, whereas the twisted-circle readout is entire in $q$ and encodes only global winding-sector rearrangements.
Although both recover the standard relative zeta-regularised one-loop contribution at $q=1$, the finite-difference family distinguishes the different local and global structures through which that limit is reached.

%%%%%%%%%%%%%%%%%%%%%%%%%%%%%%%%%%%%%%%%%%%%%%%%%%%%%%%%%%%%%%%
%%%%%%%%%%%%%%%%%%%%%%%%%%%%%%%%%%%%%%%%%%%%%%%%%%%%%%%%%%%%%%%
\section{Four-Dimensional \texorpdfstring{$\bm{a_{4}}$}{a4} Response and Full-Spectrum Information}\label{sec:4dtheory}

The relative examples separated local from global spectral information through cancellations between pairs of operator spectra.
We now examine the corresponding distinction within four-dimensional one-loop spectra, where the distinguished point $q=1$ has complementary local and full-spectrum roles.
For a four-dimensional Laplace-type operator, local invariants of mass dimension four in the action density belong to the $a_{4}$ hierarchy, whose Mellin and readout positions are $s=0$ and $q=1$, respectively, with $\chi_{4}(4,q;2)=1-q$.
At this point, the regular special value $\zeta_{A}(0)=a_{4}(A)$ determines the local scale response through Eq.~\eqref{eq:gen_scale_resp}, while the node-coalescence limit recovers the standard zeta-regularised one-loop contribution.

At generic regular values of $q$, the scale response instead depends on $\zeta_{A}(q-1)$ and hence generally retains information about masses, boundary conditions, thresholds, and the full spectrum.
We illustrate this distinction first for a constant scalar background with a single mass scale and then for a background gauge field with several gauge, matter, and ghost fluctuation sectors.
In the latter case, the separate spectral responses remain distinguishable at finite $q$ before their local contributions combine at $q=1$ into the standard renormalisation-group (RG) coefficient.

%%===================
\subsection{Single-Mass-Scale \texorpdfstring{$\bm{\phi^{4}}$}{phi4} Response}

For a real scalar field on four-dimensional Euclidean space, let
\begin{equation}
U_{\mathrm{cl}}(\phi)=\frac{1}{2}m^{2}\phi^{2}+\frac{\lambda}{4!}\phi^{4},\quad m^{2}>0,
\end{equation}
and consider the quadratic fluctuation operator about a constant background,
\begin{equation}
\widetilde{A}_{\phi}=-\partial^{2}+M^{2}(\phi),\quad M^{2}(\phi)=m^{2}+\frac{\lambda}{2}\phi^{2},\quad A_{\phi}=\frac{\widetilde{A}_{\phi}}{\mu^{2}}.
\end{equation}
We formally expand $\zeta_{A_{\phi}}(s)$ about $\phi=0$ and denote by $\mathcal{P}_{\phi^{4}}$ the projection onto the term proportional to $\phi^{4}$.
The $a_{0}$ contribution is independent of the background field, while the $a_{2}$ contribution is at most quadratic in $\phi$.
The first term surviving this projection therefore belongs to the $a_{4}$ hierarchy.
One obtains
\begin{equation}
\mathcal{P}_{\phi^{4}}\biggl[\frac{\zeta_{A_{\phi}}(s)}{V}\biggr]=\frac{\lambda^{2}}{128\pi^{2}}\left(\frac{\mu^{2}}{m^{2}}\right)^{s}\phi^{4}.
\end{equation}
The corresponding finite-difference one-loop readout is
\begin{equation}
\mathcal{P}_{\phi^{4}}\bigl[U_{q}^{(1)}(\phi)\bigr]=\frac{1}{2(1-q)}\mathcal{P}_{\phi^{4}}\left[\frac{\zeta_{A_{\phi}}(q-1)-\zeta_{A_{\phi}}(0)}{V}\right]=-\frac{\lambda^{2}}{128\pi^{2}}F_{2-q}\left(\frac{\mu^{2}}{m^{2}}\right)\phi^{4}.
\end{equation}
This expression embeds the standard logarithmic correction in a one-parameter family of mass-dependent profiles.
For $q\neq 1$ at regular readout positions, the dimensionless mass ratio $\mu^{2}/m^{2}$ enters through a power law, so that the sensitivity of the readout to the mass scale $m$ changes continuously with $q$.
The standard result is recovered only at the node-coalescence point:
\begin{equation}
\mathcal{P}_{\phi^{4}}\bigl[U_{1}^{(1)}(\phi)\bigr]=-\frac{\lambda^{2}}{256\pi^{2}}\left(\ln\frac{\mu^{2}}{m^{2}}\right)\phi^{4}.
\end{equation}
This is the usual logarithmic one-loop term.
The explicit scale response, with $m$, $\phi$, and $\lambda$ held fixed, is
\begin{equation}
\mathcal{P}_{\phi^{4}}\biggl[\mu\frac{\partial U_{q}^{(1)}}{\partial\mu}\biggr]=-\frac{\lambda^{2}}{128\pi^{2}}\left(\frac{\mu^{2}}{m^{2}}\right)^{q-1}\phi^{4}.
\end{equation}
At $q=1$, the mass-ratio dependence disappears.
Requiring cancellation between the scale dependence of the classical term $\lambda\phi^{4}/4!$ and that of the one-loop term gives $\beta_{\lambda}^{(1)}=3\lambda^{2}/(16\pi^{2})$.
This is the standard one-loop $\beta$ function \cite{Collins84,ZinnJustin02}.

The finite-$q$ result contains information not retained by this local RG coefficient.
Whereas $\beta_{\lambda}^{(1)}$ records only the logarithmic scale response at $q=1$, the factor $\bigl(\mu^{2}/m^{2}\bigr)^{q-1}$ preserves the finite dependence on the mass scale.
The parameter $q$ therefore provides an auxiliary coordinate for comparing the local universal response with the finite mass-sensitive response of the same fluctuation operator.
It does not define a new $\beta$ function, but prevents this mass dependence from being collapsed prematurely into its logarithmic value at the node-coalescence point.

%%===================
\subsection{Gauge-Sector Readout and Local RG Data}

We next consider a background gauge field with several fluctuation sectors.
Gauge, matter, and ghost operators combine at $q=1$ into the same local RG coefficient but remain spectrally distinct at generic regular values of $q$, making gauge theory a natural setting for the sector-resolved readout.

After background-covariant gauge fixing, let the dimensionful quadratic fluctuation operators be of Laplace type,
\begin{equation}
\widetilde{A}_{i}=-D_{i}^{2}+E_{i},\quad A_{i}=\frac{\widetilde{A}_{i}}{\mu^{2}}.
\end{equation}
Here, $D_{i}$ and $E_{i}$ are respectively the background-covariant derivative and endomorphism on the field bundle of the $i$th sector.
Writing $\Omega_{i,\mu\nu}=[D_{i,\mu},D_{i,\nu}]$, we denote its gauge component in that representation by $F_{i,\mu\nu}$.
The representation dependence of the full spectrum is carried by $D_{i}$ and $E_{i}$, while the corresponding local heat-kernel coefficients depend on $\Omega_{i,\mu\nu}$ and $E_{i}$.
We assume that the fluctuation operators entering the zeta functions are strictly positive and self-adjoint; zero-mode collective-coordinate factors, when present, require separate treatment.
Including statistical signs and field multiplicities in coefficients $c_{i}$, we define the total spectral zeta function by
\begin{equation}
\mathcal{Z}^{(1)}(s)=\sum_{i}c_{i}\zeta_{A_{i}}(s)
\end{equation}
and the corresponding finite-difference one-loop readout by
\begin{equation}
\mathcal{W}_{q}^{(1)}=\frac{1}{2}\frac{\mathcal{Z}^{(1)}(q-1)-\mathcal{Z}^{(1)}(0)}{1-q}.
\end{equation}
Throughout this subsection, explicit scale derivatives of the readout are taken at fixed dimensionful fluctuation operators $\{\widetilde{A}_{i}\}$.
Under this convention,
\begin{equation}\label{eq:4d_exact_scale_resp}
\mu\frac{\partial\mathcal{W}_{q}^{(1)}}{\partial\mu}=-\mathcal{Z}^{(1)}(q-1)=-\sum_{i}c_{i}\zeta_{A_{i}}(q-1).
\end{equation}
At a generic regular value of $q$, the functions $\zeta_{A_{i}}(q-1)$ remain associated with the individual fluctuation sectors, so changing $q$ modifies their relative Mellin weighting without altering the operators or their spectra.

For a four-dimensional Laplace-type operator, the local gauge kinetic term arises from the $E_{i}^{2}$ and $\Omega_{i,\mu\nu}\Omega_{i}^{\mu\nu}$ contributions to $a_{4}(A_{i})$.
We define their summed $F^{2}$ projection by
\begin{equation}
\mathcal{P}_{F^{2}}\bigl[\mathcal{Z}^{(1)}(0)\bigr]=\mathcal{C}_{F}\int\dd^{4}x\,\tr\bigl(F_{\mu\nu}F^{\mu\nu}\bigr),
\end{equation}
where $\mathcal{P}_{F^{2}}$ extracts the term quadratic in the background field strength, and $\mathcal{C}_{F}$ includes the corresponding sector multiplicities and statistical signs through the coefficients $c_{i}$ \cite{Gilkey95,Vassilevich03,Kirsten01}.
Taking the node-coalescence limit in Eq.~\eqref{eq:4d_exact_scale_resp} gives
\begin{equation}
\mathcal{P}_{F^{2}}\left[\lim_{q\to 1}\mu\frac{\partial\mathcal{W}_{q}^{(1)}}{\partial\mu}\right]=-\mathcal{C}_{F}\int\dd^{4}x\,\tr\bigl(F_{\mu\nu}F^{\mu\nu}\bigr).
\end{equation}
Thus, at the distinguished point $q=1$, the summed $F^{2}$ scale response reduces to the single local coefficient $\mathcal{C}_{F}$.

With the classical normalisation
\begin{equation}
S_{\mathrm{YM}}=\frac{1}{4g_{\mathrm{YM}}^{2}}\int\dd^{4}x\,\tr\bigl(F_{\mu\nu}F^{\mu\nu}\bigr),
\end{equation}
where $F_{\mu\nu}$ is defined without an explicit factor of $g_{\mathrm{YM}}$, the cancellation of the explicit one-loop scale dependence against the implicit scale dependence of the classical coupling gives $\beta(g_{\mathrm{YM}})=-2\mathcal{C}_{F}g_{\mathrm{YM}}^{3}$.
This is the standard one-loop relation in the stated normalisation \cite{tHooft72,Collins84}.
For $q\neq 1$, the sector-resolved functions $\zeta_{A_{i}}(q-1)$ retain their distinct masses, threshold scales, background couplings, and spectral distributions, but do not define a new beta function, Wilsonian decoupling, or threshold matching.

The behaviour near $q=1$ also shows how the local scale response is connected to the full one-loop action.
Differentiating the scale response with respect to the readout coordinate and then taking the node-coalescence limit gives
\begin{equation}\label{eq:4d_q_derivative}
\lim_{q\to 1}\frac{\partial}{\partial q}\left[\mu\frac{\partial\mathcal{W}_{q}^{(1)}}{\partial\mu}\right]=-\mathcal{Z}^{(1)\prime}(0)=2\mathcal{W}_{1}^{(1)}.
\end{equation}
Thus, the value of the scale response at $q=1$ is local and determined by $\mathcal{Z}^{(1)}(0)$, whereas its first $q$ derivative determines the standard zeta-regularised one-loop effective action and the finite full-spectrum information absent from the local coefficient alone.

Using the nonlocal background-covariant heat-kernel expansion, the $F^{2}$ projection of the scale response at a regular value of $q$ may be written schematically as \cite{Barvinsky90,Codello13}
\begin{equation}
\mathcal{P}_{F^{2}}\left[\mu\frac{\partial\mathcal{W}_{q}^{(1)}}{\partial\mu}\right]=\sum_{i}c_{i}\int\dd^{4}x\,\tr_{i}\left[F_{i,\mu\nu}\mathcal{F}_{i,q}\biggl(-\frac{D_{i}^{2}}{\mu^{2}};\frac{m_{i}^{2}}{\mu^{2}}\biggr)F_{i}^{\mu\nu}\right].
\end{equation}
Here, $\mathcal{F}_{i,q}$ denotes the form factor obtained from the $i$th sector at the Mellin point $s=q-1$, and $m_{i}$ denotes a sector mass when present.
In the node-coalescence limit, the projected sector contributions reduce to the local $a_{4}$ coefficients whose sum gives $-\mathcal{C}_{F}$.
By Eq.~\eqref{eq:4d_q_derivative}, the corresponding first $q$ derivatives in this limit determine the nonlocal $F^{2}$ form factors of the standard one-loop effective action, including logarithmic structures such as $\ln\bigl(-D_{i}^{2}/\mu^{2}\bigr)$ in massless sectors, up to cancellations and local scheme-dependent terms.
Thus, the $q$ dependence connects the local gauge coefficient at $q=1$ to the sector-resolved full-spectrum form-factor information encoded in its neighbourhood, without introducing new effective dynamics.

Background gauge transformations that preserve the operator domains and boundary conditions act on the operators $A_{i}$ by unitary conjugation.
Their spectral zeta functions, and hence $\mathcal{W}_{q}^{(1)}$ and its $q$ derivatives, are invariant under this conjugation at every regular value of $q$, including the node-coalescence limit.
This spectral invariance alone does not establish BRST consistency, gauge-fixing-parameter independence, or anomaly cancellation, which concern the complete quantum theory.

%%%%%%%%%%%%%%%%%%%%%%%%%%%%%%%%%%%%%%%%%%%%%%%%%%%%%%%%%%%%%%%
%%%%%%%%%%%%%%%%%%%%%%%%%%%%%%%%%%%%%%%%%%%%%%%%%%%%%%%%%%%%%%%
\section{Summary and Outlook}\label{sec:conclusion}

We have formulated a finite-difference zeta readout that embeds the standard logarithmic one-loop determinant within a one-parameter meromorphic family.
The parameter $q$ simultaneously specifies the Mellin evaluation point and the spectral weight governing the $q$-dependent part of the variational response, while the node-coalescence limit $q\to 1$ recovers the standard zeta-regularised one-loop effective action.
Its contribution is to organise the information contained in the spectral zeta function along a common Mellin coordinate on which local heat-kernel data, full-spectrum information, and variational response can be compared directly.

The hierarchy-dependent indices introduced above locate the readout relative to the fixed heat-kernel hierarchy and, under the stated spectral-asymptotic assumptions, classify the corresponding high-energy behaviour.
Their variational counterparts further show that spectral sensitivity depends not only on the readout position but also on the effective high-energy order of the perturbation.
The finite-difference family therefore separates structures that are compressed into a single logarithmic determinant in the conventional formulation.

The relative examples make this distinction explicit.
For the reflectionless soliton, the first surviving local coefficient produces a genuine pole at $q=1/2$, while the general family also makes explicit the increasing infrared sensitivity as the bound state approaches zero.
For the twisted circle, the entire local power-law hierarchy cancels, and varying $q$ only redistributes the global winding-sector contributions.
The determinant value at $q=1$ alone does not reveal this difference in analytic structure.

In four dimensions, the distinguished point $q=1$ selects the local $a_{4}$ scale response and reproduces the standard one-loop RG information.
The first $q$ derivative at that point determines the full zeta-regularised one-loop effective action, including its nonlocal spectral content.
At generic regular values of $q$, the readout retains sector-resolved and mass-sensitive information that is compressed into a single local coefficient at $q=1$.
The finite-difference family therefore supplies a Mellin coordinate for comparing fixed spectra while remaining conceptually distinct from RG flow and decoupling.

Viewed in this way, the readout parameter acts as a Mellin-space prism: without altering the operator or its spectrum, varying $q$ redistributes spectral weight and renders local UV hierarchies, finite mass-sensitive contributions, infrared thresholds, and global spectral rearrangements distinguishable along a common coordinate.

A natural direction for future work is to apply the readout family to observables whose kernels select non-logarithmic spectral weights.
Possible settings include anomalous diffusion, fractional kinetics, nonlocal transport, and geometries with scale-dependent spectral dimension or crossovers between distinct Weyl regimes.
In such applications, the kernel defining the observable would supply the physical interpretation of the readout parameter $q$.

Relative spectral problems provide a second direct avenue.
Boundary conditions, interfaces, defects, holonomy, topology, bound states, and scattering data may produce partial or complete cancellation of local heat-kernel hierarchies while leaving non-trivial determinant ratios and spectral shifts.
The readout family may then distinguish responses controlled by the first surviving local hierarchy from those generated entirely by global spectral rearrangement.
Extensions to manifolds with boundary, singular or non-compact spaces, and elliptic pseudodifferential operators will require the corresponding generalised short-time expansions.

Background-field theories with several physical scales form another promising class.
When several masses, compactification scales, or symmetry-breaking scales are present, the readout may help compare the local coefficient selected at $q=1$ with the distinct finite spectral contributions retained at generic regular values of $q$.
This may be particularly useful when several fluctuation sectors contribute to the same local RG coefficient but remain spectrally distinguishable away from the node-coalescence point.

The readout coordinate and Wilsonian coarse graining describe independent operations.
A coarse-graining scale changes the effective operator and its spectrum, whereas $q$ changes only how that spectrum is interrogated.
A two-parameter construction involving both scales could therefore separate spectral evolution generated by the flow from changes in the readout applied to the evolving spectrum.
Developing such a framework would require the simultaneous flow of the effective action, its vertices, and its fluctuation operators, and lies beyond the fixed-operator one-loop setting considered here.

Beyond one loop, a single spectral variable is generally insufficient because multiple propagators, interaction vertices, and eigenfunction overlaps enter the amplitudes.
A possible extension would involve several Mellin variables associated with the spectral arguments of the propagators appearing in a loop diagram.
Whether such a multi-spectral construction admits useful analogues of the hierarchy-dependent indices introduced here remains an open question.

%%%%%%%%%%%%%%%%%%%%%%%%%%%%%%%%%%%%%%%%%%%%%%%%%%%%%%%%%%%%%%%
\begin{acknowledgments}
%%%%%%%%%%%%%%%%%%%%%%%%%%%%%%%%%%%%%%%%%%%%%%%%%%%%%%%%%%%%%%%
The author thanks Ryo Suzuki for valuable comments on an earlier version of this manuscript.
This research received no specific grant from any funding agency in the public, commercial, or not-for-profit sectors.
The views expressed herein are solely those of the author and should not be interpreted as necessarily reflecting the official policies, positions, or endorsements, whether expressed or implied, of any organisation with which the author is currently or has previously been affiliated.
%%%%%%%%%%%%%%%%%%%%%%%%%%%%%%%%%%%%%%%%%%%%%%%%%%%%%%%%%%%%%%%
\end{acknowledgments}

%%%%%%%%%%%%%%%%%%%%%%%%%%%%%%%%%%%%%%%%%%%%
%                                          %
%               APPENDICES                 %
%                                          %
%%%%%%%%%%%%%%%%%%%%%%%%%%%%%%%%%%%%%%%%%%%%

\appendix
\vspace{1.5em}
%%%%%%%%%%%%%%%%%%%%%%%%%%%%%%%%%%%%%%%%%%%%%%%%%%%%%%%%%%%%%%%
%%%%%%%%%%%%%%%%%%%%%%%%%%%%%%%%%%%%%%%%%%%%%%%%%%%%%%%%%%%%%%%
\section{Spectral Readout and Wilsonian Coarse Graining}\label{sec:spectral_operations}

This appendix distinguishes three operations that may enter spectral calculations: representing the spectrum of a fixed operator, applying a readout to that spectrum, and deforming the operator itself.
Although these operations may appear in the same calculation, they act at different levels.

A spectral representation changes only the form in which information about a fixed operator is expressed.
Eigenvalue sums, spectral measures, heat traces, resolvents, Mellin transforms, and momentum-space integrals may provide equivalent representations under appropriate analytic conditions.
Neither the operator nor its spectrum is changed.

A spectral readout instead specifies which functional is extracted from the fixed spectrum.
Spectral cutoff regularisation truncates or suppresses contributions according to their eigenvalues, the standard zeta determinant uses the derivative of the analytically continued spectral zeta function at $s=0$, and the finite-difference readout compares its values at $s=0$ and $s=q-1$.
These operations change the weighting or evaluation rule applied to the spectrum without altering the underlying operator, its domain, or its eigenvalues.
The introduction of a cutoff scale or readout parameter alone therefore does not make an operation Wilsonian.

Wilsonian coarse graining acts at a different level.
In an interacting theory, integrating out modes changes the effective action, its couplings and vertices, and generally the quadratic fluctuation operator about a given background \cite{Wilson74,Polchinski84}.
The effective operator $A_{\Lambda}$ and its spectral measure $\nu_{A_{\Lambda}}$ therefore depend on the coarse-graining scale $\Lambda$.

The independence of spectral readout and operator deformation can be expressed by combining them within a two-parameter family.
For a chosen background, let $A_{\Lambda}$ denote the quadratic fluctuation operator obtained as the Hessian of the scale-dependent action at the coarse-graining scale $\Lambda$.
Whenever $A_{\Lambda}$ belongs to the class of elliptic operators considered in this paper, applying the finite-difference readout gives
\begin{equation}
\mathcal{W}_{\Lambda,q}\equiv\mathcal{W}_{q}[A_{\Lambda}]=\frac{1}{2}\frac{\zeta_{A_{\Lambda}}(q-1)-\zeta_{A_{\Lambda}}(0)}{1-q},\quad q\neq 1.
\end{equation}
Here, $\Lambda$ labels the deformation of the effective operator, whereas $q$ labels the readout applied to the spectrum at that scale; at $q=1$, the latter is defined by the corresponding node-coalescence limit.
The two parameters are therefore complementary rather than interchangeable.
The family $\mathcal{W}_{\Lambda,q}$ may help separate changes in the spectrum generated by coarse graining from changes in the Mellin readout applied to that spectrum.
It does not, however, determine the complete Wilsonian evolution of an interacting theory, which also requires the flow of the effective action, its vertices, and the associated fluctuation operators.

%%%%%%%%%%%%%%%%%%%%%%%%%%%%%%%%%%%%%%%%%%%%%%%%%%%%%%%%%%%%%%%
%%%%%%%%%%%%%%%%%%%%%%%%%%%%%%%%%%%%%%%%%%%%%%%%%%%%%%%%%%%%%%%
\section{Finite-Dimensional \texorpdfstring{$\bm{q}$}{q}-Logarithmic Prototype}\label{sec:q-alg}

This appendix presents the finite-dimensional prototype of the finite-difference zeta readout introduced in Section~\ref{sec:readout}.
Because no analytic continuation is required, both the readout and its response weight $\lambda^{-q}$ follow directly from the matrix spectrum.

%%===================
\subsection{Finite-Dimensional Representation and Variational Response}

Consider a positive-definite Hermitian matrix $A\in\mathrm{Mat}_{N}(\mathbb{C})$ with eigenvalues $\lambda_{1},\ldots,\lambda_{N}>0$.
The finite-dimensional spectral zeta function is defined as the entire function
\begin{equation}
\zeta_{A}(s)=\Tr A^{-s}=\sum_{k=1}^{N}\lambda_{k}^{-s},
\end{equation}
with $\zeta_{A}(0)=N$.
Accordingly, the finite-difference readout may be written as
\begin{equation}\label{eq:finite_dim_readout}
D_{q}^{(N)}[A]=\frac{\zeta_{A}(q-1)-\zeta_{A}(0)}{1-q}=\frac{\Tr A^{1-q}-N}{1-q}.
\end{equation}
Under the convention adopted in the main text, the corresponding one-loop readout is $\mathcal{W}_{q}^{(N)}[A]=\frac{1}{2}D_{q}^{(N)}[A]$.
The per-eigenvalue aggregation function in Eq.~\eqref{eq:lam_aggr} defines the $q$-logarithm \cite{Tsallis88}
\begin{equation}
\ln_{q}x\equiv2F_{q}(x)=\frac{x^{1-q}-1}{1-q},\quad x>0.
\end{equation}
Here, this functional form is used only as the algebra underlying the finite-dimensional readout; no probabilistic or thermodynamic structure is assigned to the matrix spectrum.
Using the functional calculus for positive-definite matrices,
\begin{equation}
D_{q}^{(N)}[A]=\Tr\ln_{q}A=\sum_{k=1}^{N}\ln_{q}\lambda_{k}
\end{equation}
holds exactly.
Thus, the finite-dimensional readout is obtained by applying the $q$-logarithm to each eigenvalue and summing the results with the ordinary trace.
In the limit $q\to 1$,
\begin{equation}
\lim_{q\to 1}D_{q}^{(N)}[A]=\Tr\ln A=\ln\det A.
\end{equation}

Now consider a smooth family of positive-definite Hermitian matrices $A(\tau)$.
No commutativity between $A$ and $\delta A$ is assumed.
The trace differentiation formula $\delta\Tr f(A)=\Tr\bigl(f'(A)\delta A\bigr)$ gives
\begin{equation}
\delta D_{q}^{(N)}[A]=\Tr\bigl(A^{-q}\delta A\bigr).
\end{equation}
Since $\zeta_{A}(0)=N$ is constant along the family, $\delta\zeta_{A}(0)=0$, and this expression is the complete finite-dimensional variation.
In terms of eigenvalue variations,
\begin{equation}
\delta D_{q}^{(N)}[A]=\sum_{k=1}^{N}\lambda_{k}^{-q}\delta\lambda_{k}.
\end{equation}
Thus, the response weight $\lambda^{-q}$ arises directly as the gradient of the finite-dimensional $q$-logarithmic aggregation, without invoking analytic continuation.

%%===================
\subsection{Relative Finite-Dimensional Readout}

For two positive-definite Hermitian matrices $A$ and $A_{0}$ of the same dimension $N$, we define
\begin{equation}
\Delta D_{q}^{(N)}[A;A_{0}]=D_{q}^{(N)}[A]-D_{q}^{(N)}[A_{0}].
\end{equation}
From Eq.~\eqref{eq:finite_dim_readout}, one obtains
\begin{equation}\label{eq:finite_rel_q}
\Delta D_{q}^{(N)}[A;A_{0}]=\frac{\Tr A^{1-q}-\Tr A_{0}^{1-q}}{1-q}.
\end{equation}
The corresponding one-loop relative readout is $\Delta\mathcal{W}_{q}^{(N)}[A;A_{0}]=\frac{1}{2}\Delta D_{q}^{(N)}[A;A_{0}]$.
Since the two matrices have the same dimension, the reference term $N$ contained in the absolute readout cancels in the relative difference.
In the limit $q\to 1$,
\begin{equation}
\lim_{q\to 1}\Delta D_{q}^{(N)}[A;A_{0}]=\ln\frac{\det A}{\det A_{0}}.
\end{equation}
Equation~\eqref{eq:finite_rel_q} is therefore the finite-dimensional prototype of the relative zeta readout used in the main text, with cancellation of the common reference term $N$ providing its basic algebraic mechanism.
Positive definiteness excludes zero modes, while finite dimensionality removes the need to address short-time singular asymptotics, analytic continuation, continuous spectra, and threshold states.
These additional analytic issues arise in the corresponding infinite-dimensional problems.

%%%%%%%%%%%%%%%%%%%%%%%%%%%%%%%%%%%%%%%%%%%%%%%%%%%%%%%%%%%%%%%
%%%%%%%%%%%%%%%%%%%%%%%%%%%%%%%%%%%%%%%%%%%%%%%%%%%%%%%%%%%%%%%
\section{Solvable Relative Models}\label{sec:relative_examples_details}

This appendix derives the spectral-shift and winding representations, analytic continuations, and special-function formulae used for the two relative models in Section~\ref{subsec:rel_models}.

%%===================
\subsection{Scattering Representation for the Reflectionless Soliton}

For the operator pair in Eq.~\eqref{eq:soliton}, the sign convention of Eq.~\eqref{eq:rel_zeta_spect_shift} gives a spectral shift function with a downward unit jump at the bound-state eigenvalue and a continuous part determined by the scattering phase.
Writing
\begin{equation}
\lambda_{\mathrm{b}}=\frac{M^{2}-\kappa^{2}}{\mu^{2}},\quad\lambda_{\mathrm{th}}=\frac{M^{2}}{\mu^{2}},
\end{equation}
its explicit form is
\begin{equation}\label{eq:sol_spectral_shift}
\xi_{\mathrm{sol}}(\lambda)=
\left\{
\begin{array}{c@{\qquad}r@{}l}
\hspace{0.64em}0, & 0\leq{} & \lambda<\lambda_{\mathrm{b}},\\[0.3em]
-1, & \lambda_{\mathrm{b}}\leq{} & \lambda<\lambda_{\mathrm{th}},\\[0.3em]
-\displaystyle\frac{1}{\pi}\delta\bigl(\sqrt{\mu^{2}\lambda-M^{2}}\bigr), && \lambda\geq\lambda_{\mathrm{th}}.
\end{array}
\right.
\end{equation}
Substitution into Eq.~\eqref{eq:rel_zeta_spect_shift} gives Eq.~\eqref{eq:sol_rel_zeta_start}.
Its continuum integrand behaves as $k^{-2s-2}$ for $k\to\infty$, so the integral converges directly for $\Re s>-1/2$.
Evaluating it by the substitution $k=M\sqrt{u/(1-u)}$ gives
\begin{equation}
\int_{0}^{\infty}\frac{\bigl(k^{2}+M^{2}\bigr)^{-s}}{k^{2}+\kappa^{2}}\dd k=\frac{\sqrt{\pi}}{2\kappa}\frac{\Gamma\bigl(s+\frac{1}{2}\bigr)}{\Gamma(s+1)}M^{-2s}{}_{2}\mathrm{F}_{1}\!\left(s,\frac{1}{2};s+1;1-\frac{\kappa^{2}}{M^{2}}\right),
\end{equation}
and hence the meromorphic continuation
\begin{equation}\label{eq:sol_rel_zeta_closed}
\Delta\zeta_{\mathrm{sol}}(s)=\left(\frac{M^{2}-\kappa^{2}}{\mu^{2}}\right)^{-s}-\frac{\Gamma\bigl(s+\frac{1}{2}\bigr)}{\sqrt{\pi}\,\Gamma(s+1)}\left(\frac{M^{2}}{\mu^{2}}\right)^{-s}{}_{2}\mathrm{F}_{1}\!\left(s,\frac{1}{2};s+1;1-\frac{\kappa^{2}}{M^{2}}\right).
\end{equation}
Setting $s=q-1$ in Eq.~\eqref{eq:sol_rel_zeta_closed} and using Eq.~\eqref{eq:rel_q_zeta} gives Eq.~\eqref{eq:sol_rel_q}.

The bound-state--continuum decomposition also gives the relative heat trace.
Using
\begin{equation}
\int_{0}^{\infty}\frac{e^{-tk^{2}}}{k^{2}+\kappa^{2}}\dd k=\frac{\pi}{2\kappa}e^{\kappa^{2}t}\erfc\bigl(\kappa\sqrt{t}\bigr)
\end{equation}
and adding the bound-state contribution yields Eq.~\eqref{eq:rel_ker_sol}.
Its expansion through the next nonzero order is
\begin{equation}
\Delta K_{\mathrm{sol}}(t)=\frac{2\kappa}{\sqrt{\pi}}\,t^{1/2}-\frac{2\kappa}{\sqrt{\pi}}\left(M^{2}-\frac{2\kappa^{2}}{3}\right)t^{3/2}+O\bigl(t^{5/2}\bigr),\quad t\to 0^{+}.
\end{equation}
This confirms the cancellation of the $t^{-1/2}$ Weyl term and identifies the first surviving coefficient of the relative heat trace for the dimensionful pair as $\Delta\widetilde{a}_{2}=2\kappa/\sqrt{\pi}$.
At $s=0$, the bound-state contribution to the scattering representation is $+1$, while the continuum contribution is
\begin{equation}
\frac{1}{\pi}\int_{0}^{\infty}\frac{\dd\delta(k)}{\dd k}\dd k=\frac{\delta(\infty)-\delta(0)}{\pi}=-1.
\end{equation}
Thus, $\Delta\zeta_{\mathrm{sol}}(0)=0$, and the node-coalescence limit of Eq.~\eqref{eq:sol_rel_q} gives Eq.~\eqref{eq:sol_rel_det}.

%%===================
\subsection{Winding Representation for the Twisted Circle}

Starting from Eq.~\eqref{eq:twist_rel_zeta_start}, we use the Schwinger representation in the half-plane of absolute convergence and apply Poisson resummation to the shifted Gaussian sums:
\begin{equation}
\sum_{n\in\mathbb{Z}}\left[e^{-t(n+\alpha)^{2}}-e^{-tn^{2}}\right]=\sqrt{\frac{\pi}{t}}\sum_{k\in\mathbb{Z}}e^{-\pi^{2}k^{2}/t}\left(e^{2\pi i k\alpha}-1\right).
\end{equation}
The $k=0$ term vanishes identically, leaving only the nonzero winding sectors.
Combining the $k$ and $-k$ terms and using
\begin{equation}
\int_{0}^{\infty}t^{\nu-1}e^{-\beta t-\gamma/t}\dd t=2\left(\frac{\gamma}{\beta}\right)^{\nu/2}\mathrm{K}_{\nu}\bigl(2\sqrt{\beta\gamma}\bigr),\quad\Re\beta>0,\quad \Re\gamma>0,
\end{equation}
gives
\begin{equation}\label{eq:twist_rel_zeta_closed}
\Delta\zeta_{\alpha}(s)=\left(\frac{\mu L}{2\pi}\right)^{2s}\frac{4\sqrt{\pi}}{\Gamma(s)}\sum_{k=1}^{\infty}\left[\cos(2\pi k\alpha)-1\right]\left(\frac{\pi k}{c}\right)^{s-\frac{1}{2}}\mathrm{K}_{s-\frac{1}{2}}(2\pi kc),\quad c=\frac{mL}{2\pi}.
\end{equation}
For $m>0$, the modified Bessel function decays exponentially as $k\to\infty$, locally uniformly for $s$ in compact subsets of the complex plane.
The series therefore converges locally uniformly and, together with the factor $\Gamma(s)^{-1}$, defines an entire continuation of the relative zeta function.
Applying the same Poisson resummation directly to the relative heat trace gives Eq.~\eqref{eq:twist_rel_heat_compact}.
Because every surviving term contains the factor $e^{-L^{2}k^{2}/(4t)}$, its short-time behaviour satisfies Eq.~\eqref{eq:rel_ker_twist_zero}, and every coefficient in the ordinary local power-law heat-kernel expansion vanishes.
The series in Eq.~\eqref{eq:twist_rel_zeta_closed} is regular at $s=0$, while $\Gamma(s)^{-1}$ vanishes there, so $\Delta\zeta_{\alpha}(0)=0$.
Setting $s=q-1$ in Eq.~\eqref{eq:twist_rel_zeta_closed} then gives Eq.~\eqref{eq:twist_rel_q}.
Since the relative zeta function is entire, the apparent singularity of the readout at $q=1$ is removable, and its node-coalescence limit reproduces Eq.~\eqref{eq:twist_rel_det}.

%%%%%%%%%%%%%%%%%%%%%%%%%%%%%%%%%%%%%%%%%%%%
%                                          %
%               BIBLIOGRAPHY               %
%                                          %
%%%%%%%%%%%%%%%%%%%%%%%%%%%%%%%%%%%%%%%%%%%%

%\bibliographystyle{apsrev4-2}

%

\end{document}